\documentclass{caosp}
\usepackage{graphicx}
\articleNo{}
\pubyear{2008}
\volume{38}
\volnumber{2}
\firstpage{1}
\received{Jan 4, 2008}
\accepted{Jan 22, 2008}
\begin{document}
\hauthor{P.L.\,North, J.\,Babel and D.\,Erspamer}
\title{The evolution of Ap stars}
\author{
        P.L.\,North \inst{1} 
      \and 
        J.\,Babel \inst{2},
      \and 
        D.\,Erspamer \inst{3}
       }
\institute{
           Ecole Polytechnique F\'ed\'erale de Lausanne (EPFL), Observatoire,
	   CH--1290 Versoix, \email{pierre.north@epfl.ch}
         \and 
           rue des Battieux 36,
	   CH--2000 Neuch\^atel, Switzerland
         \and 
           Lyc\'ee-Coll\`ege de l'Abbaye, CH--1890 St-Maurice, Switzerland
          }
\date{March 8, 2003}
\maketitle

\begin{abstract}

The many peculiarities of Ap stars (not only chemical ones, but also
magnetic field and slow rotation) may vary during the main-sequence evolution
of these stars.  We review here briefly the evidences found in the last thirty
years for such evolution, with an emphasis on the more recent research. The
position in the HR diagram of low mass Ap stars with a significant surface
magnetic field (as measured by us) is reviewed as well.  


\end{abstract}

\section{Introduction}

We limit this review to the {\it magnetic} Bp-Ap stars in the traditional
sense, i.e.~those classified as Si or SrCrEu (hereafter ``Ap stars''). Their
characteristics are well known in broad outline, even though they may vary a
lot from one individual to the next.  Many results that were reported
in a previous review (North 1993) will only be briefly mentioned here, if at 
all.

Every characteristic of Ap stars is likely to change with time: frequency
$N(Ap)/[N(A)+N(Ap)]$, type and strength of peculiarity, rotation ($V\sin i$ and
period), strength and geometry of the magnetic field.  Any observed change --
or lack of it -- is expected to put strong constraints on theoretical models
and to shed light on the nature and origin of those peculiarities.  Ap stars
lie on the main sequence (hereafter MS), as has been known for decades, but
they may also be in the pre-MS phase, as proposed by (e.g.) Alecian et
al.~(R-Alecian, these proceedings). Then, towards the end of an evolutionary
track are the white dwarfs, some of which are strongly magnetic. It is tempting
to relate the latter to magnetic Ap stars, as did Kawka and Vennes (2004) and
Putney (1996).  But as those authors have pointed out, should the magnetic flux
be conserved, only those white dwarfs with a field $B\ga 10^7$~G can be
descendants of Ap stars.

There are two empirical ways of studying the evolution of stars
in general:

{\bf -- Through galactic open clusters:}
This method is the most secure since it allows precise ages to be determined,
using the stars lying close to the turn-off.  The distance and reddening of
open clusters are also well known, as is the bulk metallicity, which is
especially interesting since the chemical peculiarities of a single Ap star are
sufficiently bizarre to mask that particular property.  On the other hand,
membership of Ap stars in the cluster must be well attested, as must also the
completeness of their detection down to a given mass.

{\bf -- Through field Ap stars with known surface gravity or age:} This method
was employed decades ago by Wolff (1975), who used both $V\sin i$ and the
rotation period $P$ to determine $R\sin i$ and to study the evolution of the
equatorial velocity with time.  Photometric estimates of $\log g$ were used by
North (1985, 1986) for BpSi stars, with some success even though such estimates
are much less reliable for Ap stars than for normal ones: the rms error is
typically $0.3$~dex instead of $0.1$~dex, but the total range of the $\log g$
values amounts to about $0.6$~dex on the MS, so purely photometric estimates
may remain interesting when a large enough number of stars is available.  The
Hipparcos results have allowed more secure estimates of $\log g$: the
bolometric luminosity is determined from parallax and $T_{\mathrm{eff}}$, and
theoretical evolutionary tracks give access to the mass by interpolation.
However, since evolution is much slower near the Zero Age Main Sequence (ZAMS)
than near the Terminal Age Main Sequence (TAMS), in practice this method does
not provide time resolution near the ZAMS, as pointed out by North (1993).
Using age directly instead of $\log g$ leads to the same limitation (Kochukhov
and Bagnulo 2006; Landstreet et al. 2007 -- see especially their Fig.~3).  The
latter authors also underlined the additional uncertainty introduced by the
scatter of bulk metallicities in the solar neighbourhood.  One should also keep
in mind that this method relies on the assumption that an Ap star with a given
mass and metallicity follows the same evolutionary track as a normal A-type
star.  Nevertheless, it is useful for the late stages of evolution on the MS,
so it is perfectly complementary to the cluster method.

\section{The evolution of Ap stars}
We briefly review the work done and the results obtained so far regarding
the evolution of the most conspicuous characteristics of Ap stars.

\subsection{Frequency}

{\bf In the field:} Frequency is a simple concept, but there is some confusion
in the literature regarding its exact meaning. Wolff (1968) computed an overall
Ap-star ``incidence'' of about $10$\% among the bright MS stars in the interval
$-0.19 < (B-V) < +0.20$.  The sample is magnitude limited ($V<5.0$), so the
above figure can give the expected number of Ap stars that a future survey will
yield, but has no direct physical meaning.  Note also that this incidence is
defined by the ratio $R\equiv\frac{N(Ap)}{N(IV)+N(V)}$, $N(IV)$ and $N(V)$
being the respective numbers of normal A-type stars of luminosity classes $IV$
and $V$. A better choice would be to define
$R\equiv\frac{N(Ap)}{N(Ap)+N(IV)+N(V)}$.  In any case, it is not a
volume-limited sample, since at the limiting apparent magnitude B5 stars are
much more distant than A5 ones.  On the other hand, a volume-limited sample
such as that defined by Power et al.~(P26, these proceedings) on the basis of
Hipparcos parallaxes, although quite relevant in principle, is unsatisfactory
in practice since there are no Bp stars more massive than about $3.5~{\rm
M}_\odot$ within $100$~pc.

The magnitude-limited frequency $f_{\rm m}$ for a mix of Ap stars of various
masses is the mean of the true (volume limited) frequencies $f_{\rm i}$,
weighted by the total number of stars present in the respective mass bins and
corresponding volumes.  For example, a mix of ApSi stars with $M=3.5\,{\rm
M}_\odot$, $M_V=-1.2$, $f_{\rm Si}=12$\% and of ApSrCrEu stars with
$M=2.25\,{\rm M}_\odot$, $M_V=1.1$, $f_{\rm SrCrEu}=4$\% will give $f_{\rm
m}=8.2$\% under the assumption of a Salpeter initial mass function (the
limiting apparent magnitude does not matter as long as visual absorption is
neglected). The volume-limited overall frequency $f_{\rm v}$ of the same mix is
the mean weighted by the respective stellar number densities, giving $f_{\rm
v}=6.1$\%.

\noindent{\bf In Galactic open clusters:} The first studies addressed the
simple question whether Ap stars can be found at all in open clusters, and (if
so) if their frequency is the same there as in the field.  Those studies were
in general based on spectral classification (Young and Martin 1973; Hartoog
1976; van Rensbergen et al. 1978).  The very first attempt to detect Ap stars
in clusters by photometric means was made by a student of B.~Hauck, using what
became known as the $\Delta (V1-G)$ parameter of Geneva photometry (Steiger
1974).  Maitzen (1976) then devised the $\Delta a$ photometric system
specifically to detect Ap stars through their $\lambda 5200$\AA~spectral
feature, and used it extensively to take a census of Ap stars in Galactic
clusters (e.g.~Maitzen and Hensberge 1981; Maitzen and Wood 1983). The
frequency of Ap stars in clusters was explored independently using the Geneva
photometry (North and Cramer 1981). It was generally agreed that the frequency
of Ap stars is roughly the same in clusters as in the field, though the
relatively small samples did not allow a very definite conclusion.

\noindent{\bf Does the frequency of Ap stars change with age?}  That is
equivalent to asking: ``Do young Ap stars exist at all, or do the peculiarities
develop later?''.  The question was addressed, on the basis of spectral
classification, by van Rensbergen et al.~(1978) and Abt (1979) using clusters,
and by Abt and Cardona (1983) using binary stars.  The latter authors reported
an increasing trend of frequency with age, but independent studies (North and
Cramer 1981; North 1993 and references therein) did not confirm that trend. An
interesting study of the Orion and Scorpio-Centaurus associations was
undertaken by Joncas and Borra (1981) and Borra et al.~(1982) by means of the
$\Delta a$ system.  They found a deficiency of Ap stars in these young
aggregates, and deduced a slow metal enrichment of their atmospheres.  On the
other hand, they also noticed a few He-weak stars in the Orion association and
suggested that those might become BpSi stars through future evolution.  That
was the first time that the idea that Ap stars could change their peculiarity
type as they evolve on the MS was raised.  The question has since
been addressed by North (1993) on the basis of a compilation of the literature.
No clear trend was obtained (except for Am stars) as a function of age, but the
frequency of Ap stars as a function of mass was, however, obtained for the
first time.

\noindent{\bf Is the frequency of Ap stars the same in a metal-poor
environment?} This is a simple but fundamental question.  If one postulates
that radiative diffusion in the presence of a magnetic field is the prevailing
cause of the chemical peculiarities, one can expect to find ``metal-poor'' Ap
stars (in the sense of their global, not atmospheric, metal content), since
there is no `a priori' reason why that mechanism should not operate at low
metallicities.  Auri\`ere et al.~(2007) have shown that all Ap SrCrEu stars with
an approximately solar global metallicity have a large-scale magnetic field, so
it appears reasonable to expect that similar Ap stars which belong to a
metal-poor population will also be magnetic.  Therefore, finding Ap stars in a
metal-poor environment probably means finding magnetic stars.  Since the origin
of the magnetic fields remains an open question (although suggestions made in
these proceedings may point to an answer) but might depend on metallicity, it
is worth exploring the matter.

Paunzen et al.~(2005, 2006) have explored four clusters and a field in the LMC,
the metal content of which ($Z\simeq 0.08$) is about a factor of 2 smaller than
that of the Sun.  They find an overall frequency $f=2.2\pm 0.6$\%, which is at
least twice as small as the frequency of $\sim 6$\% usually quoted for the solar
vicinity.  However, the volume-limited sample of Power et al.~(2008, this
conference) gives a frequency of only 1.5\%, which is the same (to within 
about 1$\sigma$) as the frequency obtained for the LMC by Paunzen et al., whose
sample can also be considered volume limited.  A reliable comparison should
take into account the different ways of defining an Ap star (spectroscopy
versus photometry) and the corresponding sample of normal B-A stars, as well as
probable biases.

\subsection{Peculiarity type}

Analyses routinely attribute chemical abundance anomalies to radiative
diffusion in the presence of gravity, with well-substantiated reasons (Michaud
1970).  During evolution on the MS, a typical A-type star will change its
$T_{\mathrm{eff}}$ by about $27$\%, e.g.~from $T_{\mathrm{eff}}=10000$~K on the
ZAMS to $T_{\mathrm{eff}}=7300$~K on the TAMS, which undoubtedly changes the
efficiency of radiative diffusion for a given ion.  Similarly, the stellar
radius will double, leading to a $0.6$~dex decrease of the logarithmic surface
gravity.  As mentioned above, Joncas and Borra (1981) had suggested the
possibility that He-weak stars become BpSi after some evolution on the MS.  For
the {\it degree} of the peculiarity, look at Klochkova and Kopylov (1986) and
references therein.

\subsection{Evolutionary state within the MS} 

{\bf In clusters:} Hubrig and Schwan (1991) and Hubrig and Mathys (1994) tried
to answer this question using the proper motion of a few CP members of two
superclusters (Hyades and UMa).  The five Ap stars studied appeared to be at
the end of their evolution on the MS when hydrogen has just been exhausted
in the core.  Since this short phase was estimated to last for about $10$\% of
the MS lifetime, it was tempting to imagine that Ap stars represent just a
quick phase in the evolution of {\it all} A-type stars, as suggested by Oetken
(1986).  However, unevolved Ap stars are known to occur in younger clusters
(North 1993), so only a very circumstantial conspiracy of nature could make
evolved non-members out of these (probably unevolved) member CP stars.  A more
complete and thorough examination of Ap stars in clusters is under way
(Landstreet et al., C-Landstreet, these proceedings).

\noindent{\bf In the field:} Hubrig et al.~(2000) used a sample with a majority
of narrow-lined and sufficiently strongly magnetic stars, in which the surface
field could be estimated from spectra observed in unpolarized light. The sample
was therefore biased towards slow rotators.  The main result was a slight
clustering of Ap stars in the middle of the MS strip, as if they appeared as
such only after having spent about $30$\% of their lifetime, and then
disappeared (i.e.~became normal A stars) slightly before they reach the core
hydrogen-burning phase.  The paper became controversial because of its
far-reaching implications regarding the origin of magnetic fields.  Hubrig et
al.~(2007) rediscussed that point on the basis of a much larger and less biased
sample (90 stars with well-defined magnetic curves); they concluded that Ap
stars with $M < 3 ~{\rm M}_\odot$ are more evolved than more massive ones.
Kochukhov and Bagnulo (2006) used an even larger sample (literature
plus recent FORS1/VLT spectra from the ESO archives), though
many stars had only one magnetic measurement. They conclude that Ap stars are,
in general, distributed uniformly in age, except for those less massive than
about $2~{\rm M}_\odot$.  The latter, although not completely absent from the
ZAMS, tend to be rare there.  Therefore, although Kochukhov and Bagnulo's
result contradicts the conclusion of Hubrig et al.~(2000) that Ap stars become
magnetic {\it only} after spending $\sim 30$\% of their MS lifetime, it does
confirm that the less massive of them are not distributed as expected across
the MS.

\section{Axial rotation} 

{\bf Projected rotational velocity:} Hartoog (1977), Abt (1979), Wolff (1981)
and Klochkova and Kopylov (1984) have used the $V\sin i$ of Ap stars in
clusters to test whether they undergo any braking during their life on the MS.
Klochkova and Kopylov (1986) considered the ratio of the average $V\sin i$ of
Ap stars of a given age and of the average $V\sin i$ of normal A stars of the
same age, in order to eliminate the effect of conservation of angular momentum,
which slows down rotation as a star expands.

\begin{figure}[ht!]
\centerline{\includegraphics[width=0.49\textwidth,clip=]{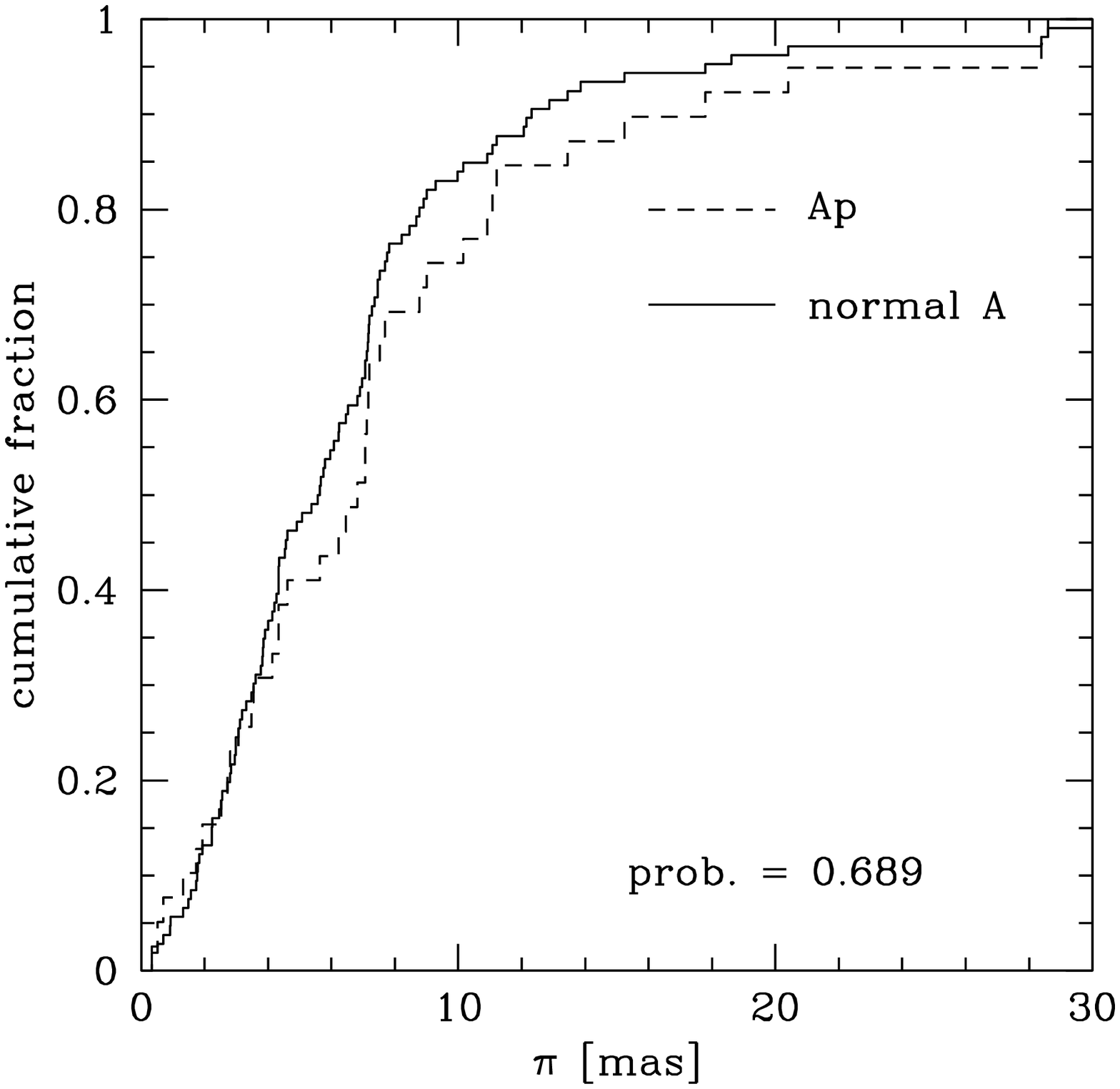} 
            \hspace{0.5cm}
            \includegraphics[width=.48\textwidth,clip=]{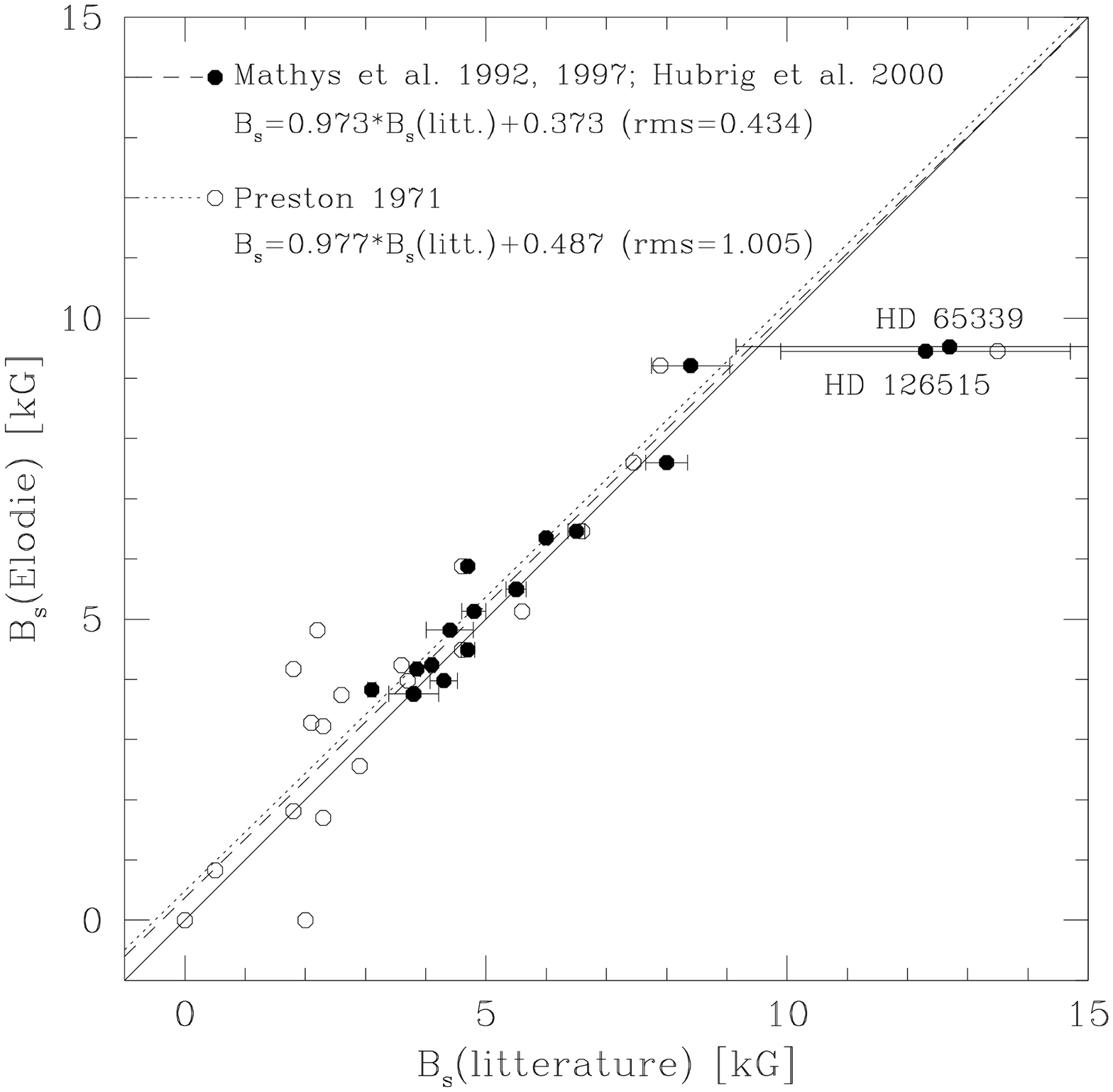}
           }
\caption{{\bf (left):} Cumulative distributions of the parallaxes of Ap stars
(broken line) and of the normal A-type stars chosen as a reference sample. 
Note the similar shapes of the distributions.
{\bf (right):} Comparison between our magnetic field estimates and those of
other authors for Ap stars in common. Very large fields deviate from the 
regression line because we applied only a simple default method involving
the measurement of the quadratic differences of the FWHMs of two 
cross-correlation functions.}
\end{figure}

\noindent{\bf Rotational period:}
When a star expands owing to its evolution, conservation of angular momentum
will cause only a slight decrease in equatorial velocity, since the latter
increases with radius, partly compensating for the conservation effect.  On the
other hand, the rotational period is a direct measure of angular velocity, so
conservation of angular momentum will imply a significant lengthening of the
rotational period.  North (1984, 1987) and Borra et al.~(1985) used the
possibility of measuring the rotational period of Ap stars through photometry
in clusters and associations, while North (1985, 1998a, b), Hubrig et al.
(2007) and Kochukhov and Bagnulo (2006) applied that idea to field Ap stars
with known ages or $\log g$. The conclusion was that for Si stars (or stars
with $M \ga 3~{\rm M}_\odot$) the period increases with age, but at the rate
expected from conservation of angular momentum if the stars are rotating more
or less like a solid body.  Thus, those stars must have acquired their slow
rotation at the time of formation; there has been no later braking.  For cooler
and less massive Ap stars, the range of periods is so wide that no conclusion
of that kind can safely be drawn.

\section{Magnetic field}

{\bf Intensity and flux:} Age will also affect the intensity of the surface
magnetic field, if the latter is a fossil of the primordial field. Borra (1981)
was the first to find a probable decrease of magnetic field strength with age,
on the basis of magnetic field measurements of Ap members of the Orion OB1
association. North and Cramer (1984) made a bold early attempt using a
correlation between photometric parameters of peculiarity and surface magnetic
field, but this correlation was later found much less significant, if any, than
first believed. Nevertheless their (fragile) conclusion, that Ap stars with
$M>3~{\rm M}_\odot$ have a field which decreases with time, was the same as
that of Borra, itself confirmed by Hubrig et al. (2007), Kochukhov and Bagnulo
(2006) and Landstreet et al. (2007). The magnetic flux (essentially $<B>R^2$
where $<B>$ is the mean value of the surface field or of a proxy for it, and
$R$ is the stellar radius) shows little change with age, if any: while
Kochukhov and Bagnulo (2006) and Hubrig et al. (2007) find it constant for
massive Ap stars ($M>3~{\rm M}_\odot$) and marginally increasing for less
massive ones ($M<3~{\rm M}_\odot$), Landstreet et al. (2007) find a decreasing
trend for massive Ap stars and a constant value for less massive ones. The
latter conclusions may be more robust than those of the other authors, because
they are based on cluster members rather than on field stars.

\noindent{\bf Geometry:}
The first attempt to study the evolution of the magnetic field geometry
provided rather uncertain conclusions, because of the large errors in the
photometric surface gravities (North 1985). Hubrig et al.~(2007) reconsidered
this matter on the basis of Hipparcos parallaxes, using the ratio $r=B_{\rm
l}^{\rm min}/B_{\rm l}^{\rm max}$ which is negative for large $\beta$ angles
between the rotational and magnetic axes, and positive for small $\beta$
angles. Interestingly, the evolution of $r$ seems quite different for massive
Ap stars compared to less massive ones.

\section{The position of cool magnetic stars on the MS}

In order to test the robustness of the conclusions of Hubrig et al.~(2000), I
chose a different sample (even though a few Ap stars will be common to both
studies), intending that both the sample and its analysis be independant.  To
take into account the weaknesses pointed out by Landstreet et al.~(2007), we
drop the Lutz-Kelker correction, and we also adopt their bolometric correction
($B.C.$).  In addition, we selected our reference sample of
normal A stars in a different way: instead of simply taking A stars closer than
$100$~pc, we selected a sample of A stars {\it having the same distribution of
parallaxes as the Ap stars}, so that any systematics linked with the parallaxes
should be the same for both samples (see Fig.~1 left). Indeed, the present sample of
Ap stars was not intended for this study: it is not a volume-limited one; it
contains Ap stars with $T_{\mathrm{eff}} < 10000$~K, $V\sin i <
20~\mathrm{km\,s^{-1}}$ and surface magnetic fields determined from echelle
spectra taken with the ELODIE spectrograph attached to the 1.93-m telescope at
Haute-Provence Observatory. The method of field determination, based on a
correlation technique, has been described by Babel et al.~(1995). Figure~1 right
shows how the surface magnetic field thus obtained correlates with other
determinations.  The typical error is a few hundreds of Gauss, so that in
practice we detected fields larger than about 1~kG for stars with $V\sin i \leq
10~\mathrm{km\,s^{-1}}$.  This sample is therefore similar to that of Hubrig et
al.~(2000) in that it contains Ap stars with a confirmed magnetic field, is
biased in favour of slow rotators, and contains about the same number of stars.

\begin{figure}[ht!]
\centerline{\includegraphics[width=0.5\textwidth,clip=]{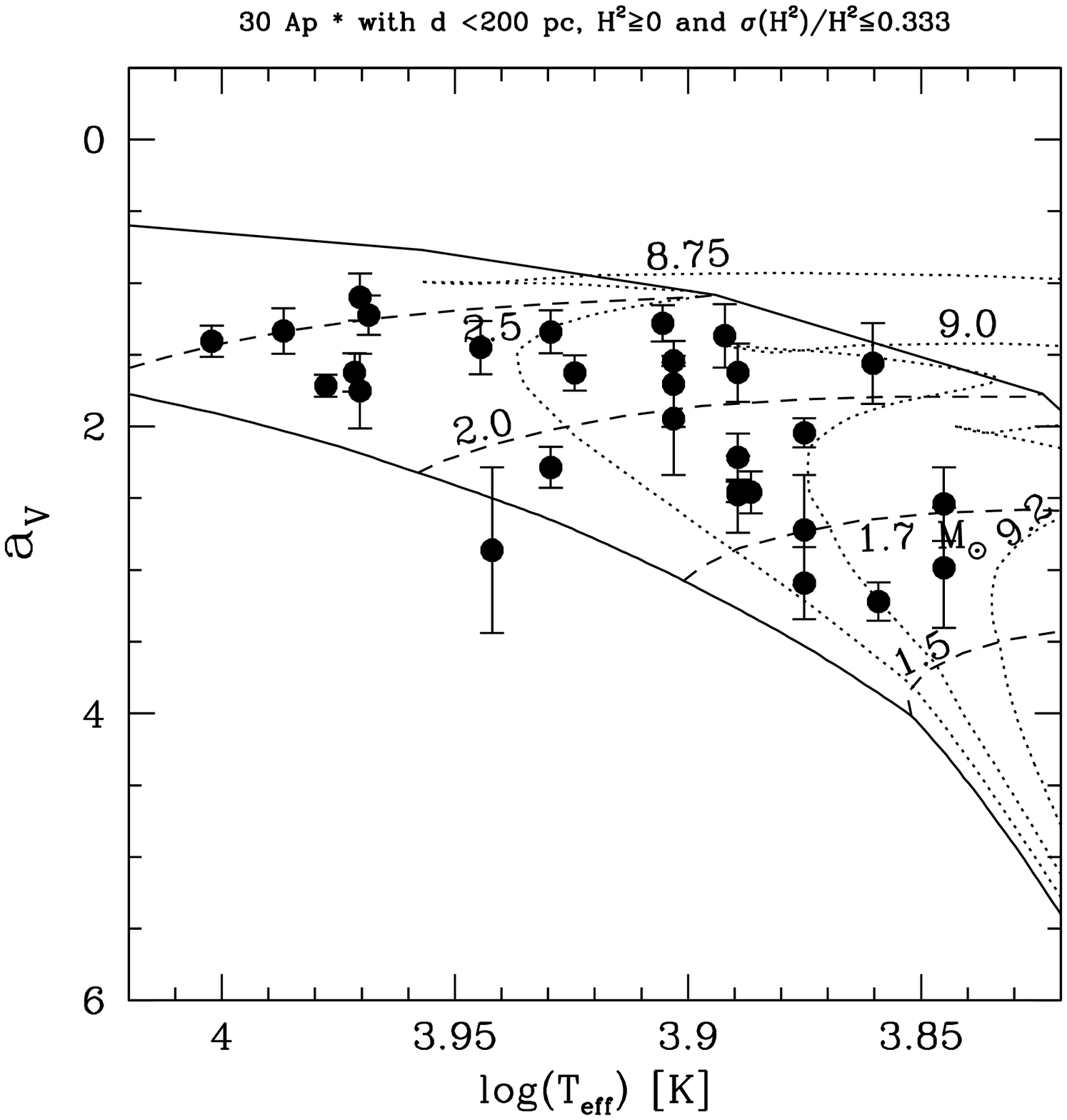} 
            \hspace{0.5cm}
            \includegraphics[width=0.5\textwidth,clip=]{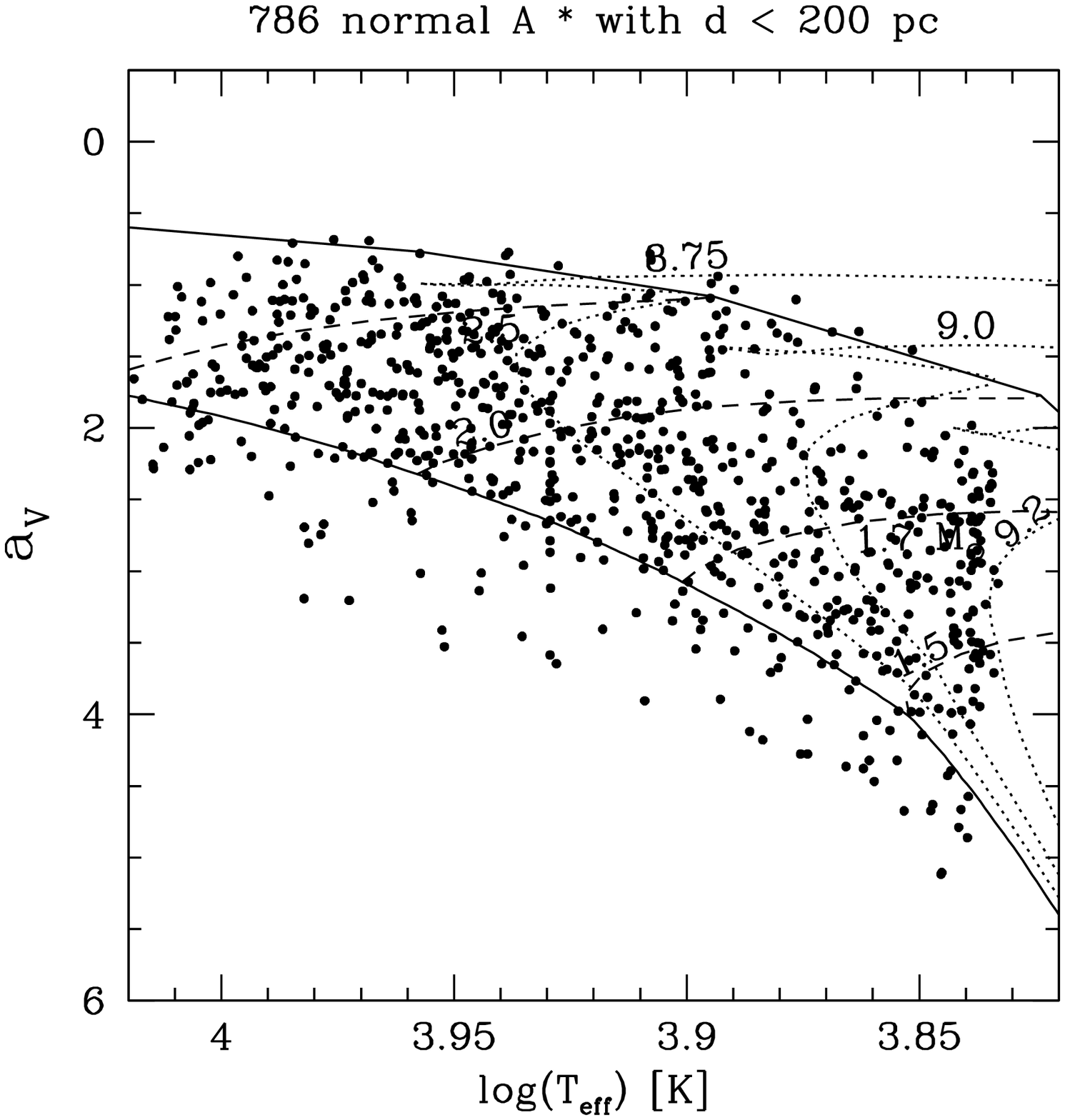}
           }
\caption{{\bf left:} Astrometric HR diagram of Ap stars of our sample
with $\pi > 5~\mathrm{mas}$. Error bars (chiefly due to parallax errors)
are symmetrical in this diagram.
{\bf right:} Same as left, but for normal A-type stars. Note that the
way the cutoff at low temperature is defined will affect the comparison between
normal A and Ap stars; here, we assume spectral types earlier than F1.}
\end{figure}
The $T_{\mathrm{eff}}$ was determined from the profile of the H$_\alpha$ line,
using Kurucz atmosphere models with $[M/H]=1.0$ and \verb+synspec+ code (Hubeny
et al.~1995).  Since the profile of this line depends not only on
$T_\mathrm{eff}$ but also on $\log g$ for $T_\mathrm{eff}\geq 8000$~K, we also
used Geneva or $uvby\beta$ photometry to remove any ambiguity.  Most of the
stars measured also have Hipparcos parallaxes which allowed us to determine
their evolutionary states.  We had to eliminate a large number of normal A
stars with $T_\mathrm{eff} < 9000$~K, which evidently outnumbered the others by
a large factor, owing to a bias in the Hipparcos Input Catalogue (had we kept
them, a very strong bias in $\log g$ would have appeared). We show the
astrometric HR diagram $a_V=\pi\cdot 10^{0.2\,M_V}=\pi\cdot
10^{0.2\,(m_V+5-A_V)}$ versus $\log T_\mathrm{eff}$ (Arenou and Luri 1999) in
Fig.~2 left and right, for Ap and normal A stars respectively.  An upper limit
to the distance of $200$~pc was set, so that both samples are volume limited
but are not necessarily complete within that volume.  At first sight, one gets
the visual impression that Ap stars tend to lie farther from the ZAMS than do
the A stars, at least for $M \leq 2\,{\rm M}_\odot$.

In order to examine the extent to which that may be significant, we built
cumulative distributions of $\log g$ and applied the two-sided
Kolmogorov-Smirnov test.  The results are shown for a limiting distance of
$200$~pc and of $500$~pc in Fig.~3 left and right respectively.  Each figure
illustrates two pairs of distributions according to stellar mass, the limit
being $2.1~{\rm M}_\odot$.  Although for stars less massive than $2.1~{\rm M}_\odot$
Fig.~3 (left) suggests a result similar to that reported by Hubrig et 
al.~(2000) (with marginal significance), Fig.~3 (right) does not show any
significant difference between the distributions of Ap and normal stars.

\begin{figure}[th!]
\centerline{\includegraphics[width=0.5\textwidth,clip=]{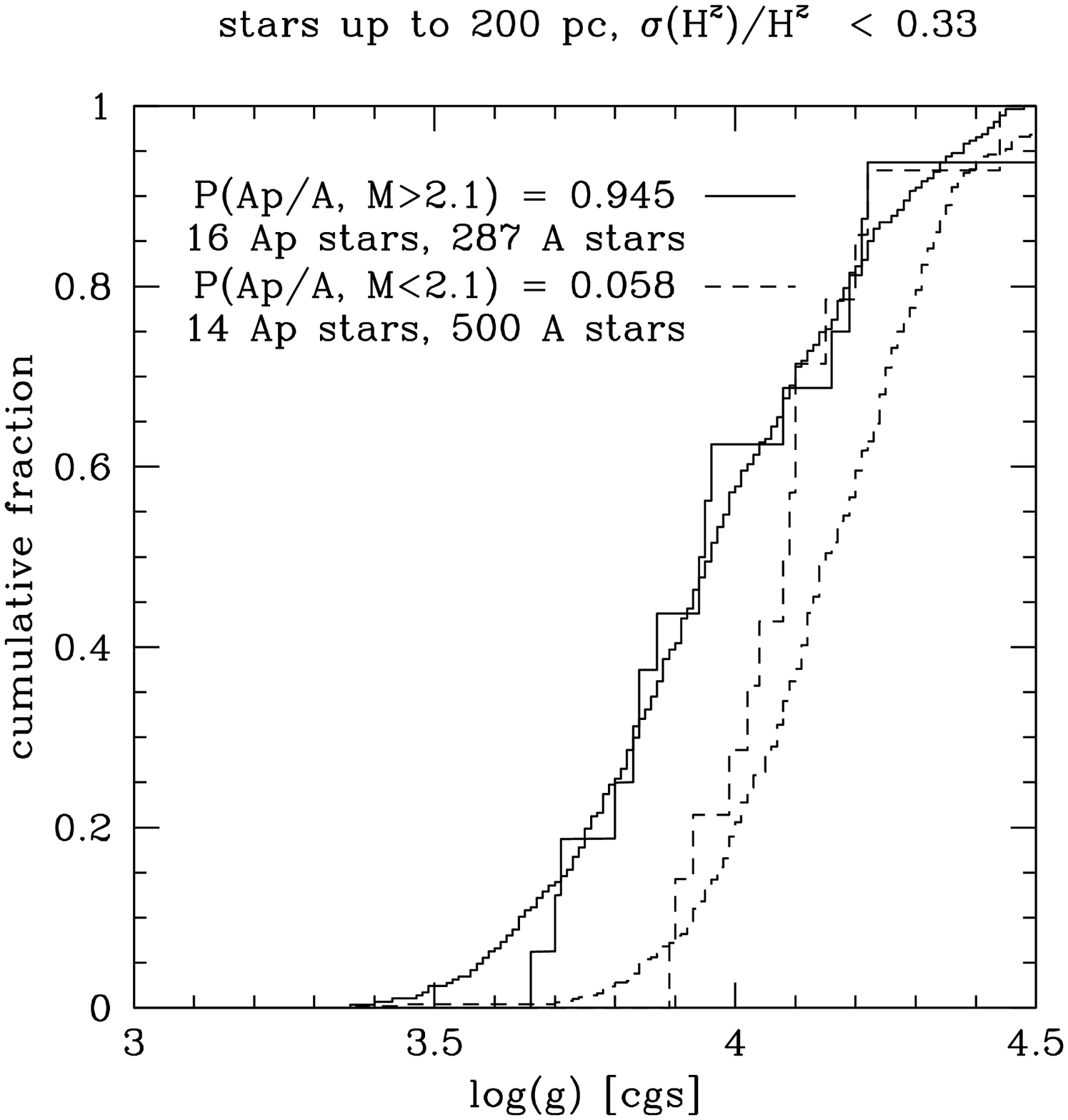} %
\hspace{0.5cm}
            \includegraphics[width=0.5\textwidth,clip=]{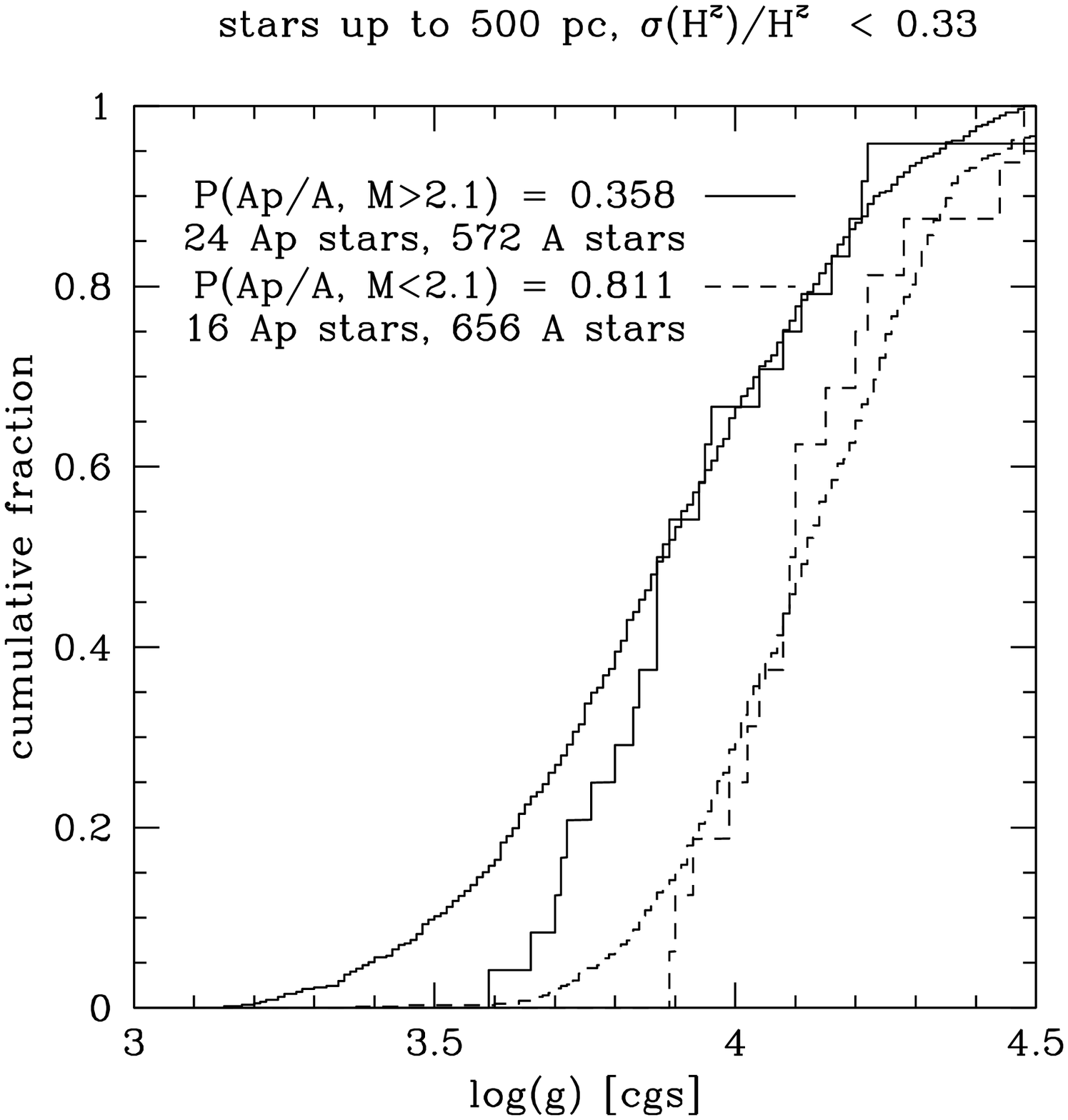}
           }
\caption{{\bf left:} Cumulative distributions of our samples of Ap and A stars
with $\pi > 5~\mathrm{mas}$.  {\bf right:} Same as left, but for stars with
$\pi > 2~\mathrm{mas}$.}
\end{figure}

\section{Conclusion}

Magnetic fields are being systematically determined for both cluster and field Ap
stars, and some evolutionary trends are emerging. Their intriguing dependence on
stellar mass is worth investigating further. However, the limited
number of Ap stars in clusters or with reliable parallaxes impedes faster
progress.  Improved Hipparcos parallaxes (van Leeuwen 2007) already increase
the sample of Ap stars with known luminosity, and the Gaia mission should
inflate it.  But a good age determination of field stars also requires reliable
$T_\mathrm{eff}$ values, which is the next limiting factor.
Formation of Ap stars is another promising topic: are magnetic Herbig Ae/Be
stars (Alecian et al., R-Alecian, these proceedings) really the progenitors of
Ap stars?  Do Ap stars form more easily in poor clusters than in rich ones
(Maitzen et al., C-Maitzen, these proceedings)? Are middle-aged Ap stars with
$M < 2~M_\odot$ really more frequent than young ones, or how can a normal A
star become Ap during its MS life?  We do not fully confirm the conclusion of
Hubrig et al.~(2000) regarding the position of Ap stars in the HR diagram.
There is no clear lack of young objects, and although Fig.~3 seems to
show a systematic lack of Ap stars with $\log g < 3.7$ ($M > 2.1~M_\odot$) and
$\log g < 3.9$ ($M < 2.1~M_\odot$), that result is not significant because of
the low sample sizes. Hubrig et al.~(2007) have cured the latter weakness to
some extent, but the incidence of the $B.C.$ uncertainty on their conclusions
should be assessed.

\acknowledgements
This work has been supported by the Swiss National Science Foundation. We thank
S. Hubrig and J. Landstreet for useful discussions.

\end{document}